# A Computationally Efficient Multiclass Time-Frequency Common Spatial Pattern Analysis on EEG Motor Imagery

Ce Zhang[1], Azim Eskandarian[2]

*Abstract*— Common spatial pattern (CSP) is a popular feature extraction method for electroencephalogram (EEG) motor imagery (MI). This study modifies the conventional CSP algorithm to improve the multi-class MI classification accuracy and ensure the computation process is efficient. The EEG MI data is gathered from the Brain-Computer Interface (BCI) Competition IV. At first, a bandpass filter and a time-frequency analysis are performed for each experiment trial. Then, the optimal EEG signals for every experiment trials are selected based on the signal energy for CSP feature extraction. In the end, the extracted features are classified by three classifiers, linear discriminant analysis (LDA), naïve Bayes (NVB), and support vector machine (SVM), in parallel for classification accuracy comparison.
The experiment results show the proposed algorithm average computation time is 37.22% less than the FBCSP (1st winner in the BCI Competition IV) and 4.98% longer than the conventional CSP method. For the classification rate, the proposed algorithm kappa value achieved 2nd highest compared with the top 3 winners in BCI Competition IV.

*Clinical Relevance*— This paper improves the motor imagery model development efficiency for future real-time motor imagery detection.

I. INTRODUCTION

EEG is one of the most popular brain signal acquisition tools because of its non-invasive and low-cost characteristics [1]. As signal processing methods and machine learning techniques improve, the applications of EEG-based BCI become more realistic and viable. Currently, EEG-based BCI prototypes are widely explored in the human motor recovery research area.

The objective of motor recovery is to rebuild disabled patients' limb function [2]. By using EEG based BCI, a patient can control artificial arms, legs, or wheelchair directly through his/her brain by conducting motor imagery [3-5]. To classify motor imagery, its spatial, frequency, and temporal domain characteristics need to be understood. In the context of spatial property, motor imagery events can be observed by brain signals in the somatosensory cortices [6]. In the context of frequency property, motor imagery initiates event-related synchronization (ERS) and event-related desynchronization (ERD) behavior for EEG signals in mu (8-13 Hz) and beta (13-30 Hz) rhythm [7]. Consequently, for the temporal property, motor imagery triggers event-related potentials (ERP) in time series EEG signals [8]. Nonetheless, these motor imagery behaviors are usually corrupted by artifacts such as eye blinking, eye movement, and head movement [9]. Therefore, a properly designed feature extraction method is necessary. Common spatial pattern (CSP) is one of the most popular feature extraction algorithms [10].

The CSP algorithm was first proposed by G. Pfurtscheller for application in motor imagery feature extraction [11]. This algorithm proved to be one of the most efficient feature extraction methods. In their study, they applied a single-trial EEG signal with two classes of motor imagery events for CSP feature extraction and achieved a maximum of 99.7% correct classification rate. Later, M. Grosse-Wentrup, et al. extended the CSP algorithm from two classes to multiple classes feature extraction by using joint approximate diagonalization (JAD) method [12].

However, since brain signals exhibit variations for every motor imagery events, the classification accuracy dramatically decreased when conducting multiple trials motor imagery analysis. Therefore, many researchers focused on improving the robustness of the CSP algorithm. K.K. Ang, et al. developed a novel algorithm, filter bank CSP (FBCSP) [13]. In their study, the raw EEG signals are bandpass-filtered into multiple frequency bands for CSP feature extraction. The classification accuracy is increased, but the computational load is high because the CSP feature extraction and feature selection algorithm has to be applied for every filtered EEG signal. Besides, some researchers combined existing signal processing skills with CSP to achieve better detection results. S. Selim and her colleagues applied a hybrid attractor metagene algorithm and a Bat optimization algorithm to CSP for a better classification rate [14]. S. Puthusserypady et al. combines adaptive filter with CSP (ACSP) to classify a 3-class motor imagery task [15]. One of their findings is that ACSP could maintain a similar detection accuracy when decreasing the size of training data. Besides traditional signal processing and machine learning methods, deep learning has also been combined with CSP. S. Sakhavi, C. Guan, and S. Yan modified the FBCSP and utilized a convolutional neural network (CNN) to classify motor imagery events [16].

[1] Ce Zhang is with Virginia Tech mechanical engineering department ASIM Lab.
[2] Azim Eskandarian is the Head of the Virginia Tech Mechanical Engineering Department and Nicholas and Rebecca Des Champs Professor, and Director of ASIM Lab.

All the methods mentioned above demonstrate that the CSP feature extraction can be improved to increase MI classification accuracy. However, compared with the traditional CSP, all current modified CSP algorithms require high computational loads, which is a challenge for real-world applications where computational resources are limited. For this reason, more efficient and simplified CSP algorithms need to be developed for multiple trials motor imagery classifications with the desired accuracy. This paper proposes a multiclass time-frequency analysis CSP algorithm, which improves the classification accuracy as compared with the conventional CSP and keeps the computation efficient as well.

Section 2 explains the open-source experiment dataset and protocol. Section 3 describes the overall proposed algorithm time-frequency analysis and the modified multiclass CSP. Finally, sections 4 and 5 present the results and conclusions, respectively.

## II. EXPERIMENT PROTOCOL

The experimental data is gathered from BCI Competition IV Dataset 2a for multiclass motor imagery classification, provided by the Institute for Knowledge Discovery and Graz University of Technology [17]. In this dataset, 22 EEG and 3 Electrooculography (EOG) channels are collected from 9 subjects through international 10-20 systems, as shown in Figure 1. During the experiment, subjects were asked to perform motor imagery on their left hand, right hand, feet, and tongue, which categorized as four classes. To collect enough data, each subject was required to perform two sessions of motor imagery and each session contained 72 trials for every class, which yield 288 trials for every subject in one session. Data from one of these two sessions were used for model training, and the other one was used for model evaluation.

Figure 2 demonstrates the experiment protocol. The EEG amplifier was sampled at 250 Hz, and a bandpass filter between 0.5 Hz to 100 Hz was applied for basic artifacts removal. In each motor imagery trial, the duration was approximately 8 seconds and started with a beep sound when a fixation cross displayed on the screen. Two seconds after the fixation cross, a cue of a motor imagery event appeared on the screen and the subject was asked to perform the corresponding motor imagery for 3 seconds. At the end of each trial, a 1-2 second break allowed the subject to prepare for the next one.

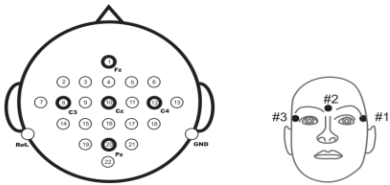

Figure 1. EEG 10-20 international system and EOG electrodes location

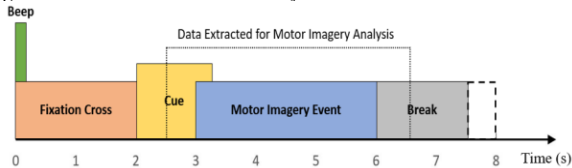

Figure 2. Experiment protocol for one trial

## III. MATERIALS & METHODS

Figure 3 presents the proposed time-frequency CSP (TFCSP) algorithm flow chart. The algorithm contains EEG preprocessing, time-frequency analysis for optimal EEG signal selection, feature extraction, and data classification. Since the EEG preprocessing procedures are standard, they are briefly discussed. Moreover, data classifiers, Linear Discriminant Analysis (LDA), Naïve Bayes (NVB), and Support Vector Machine (SVM) with radial bias function (RBF) kernel trick, are popular and common in most machine learning textbooks or papers [18, 19], which are not explained in this paper. In this section, time-frequency analysis for optimal EEG signal selection and CSP feature extraction are described in detail.

### A. Data Pre-processing

The EEG brain signal data preprocessing includes time-series data extraction and a bandpass filter. For data extraction, to cover the whole motor imagery events, data is extracted from 0.5 seconds before and 0.5 seconds after the labeled motor imagery event, as shown in the experimental protocol section (Figure 2). After time series data extraction, an $8^{th}$ degree Butterworth bandpass filter is applied to every motor imagery events. It is known that ERD/ERS behavior for motor imagery is active in mu (8-12 Hz) and beta (12-30 Hz) rhythms. Therefore, the bandpass filter bandwidth is ranged from 8 to 30 Hz.

### B. Time-Frequency Analysis

The EEG signals are non-stationary because their dynamics are varied based on different events. Hence, the traditional frequency domain transformation (Fast Fourier Transform) is not for EEG signal analysis. Thus, time-frequency analysis is introduced for non-stationary signal transformation. Currently, two methods are popular for time-frequency analysis: short-time Fourier transform (STFT) and wavelet transform (WT). Compared with the WT, the STFT conducts Fourier transform in a small uniform time window with uniform frequency band [20], as shown in Eq. 1,

$$\{x(t)\}(\tau, \omega) = \int_{-\infty}^{\infty} x(t)\omega(t-\tau)e^{-j\omega t}dt \quad (1)$$

where $\omega(t-\tau)$ is the time window function, and $x(t)$ is the time-series brain signal. Converting this equation to discrete-time STFT, it becomes:

$$\{x[n]\}(k, \omega) = \sum_{n=-\infty}^{\infty} x[n]\omega[n-k]e^{-j\omega n} \quad (2)$$

where $\omega[n-k]$ is the time window function, and $k$ is the time resolution. The benefit of the STFT analysis is that the calculation is fast and simple. The drawback of the STFT is the trade-off between the frequency resolution and time resolution. For example, to get a fine frequency resolution, the time resolution is coarse and vice versa. Since the proposed algorithm requires less stringent on the time resolution, the STFT analysis is employed for high computation efficiency.

After conducting the STFT, time-frequency results are separated into a different frequency and temporal bands, which composed a frequency and temporal band matrix, as shown in Figure 3. The matrix dimension is $m \times n$ where $m$ and $n$ represent the number of frequency band and temporal band, respectively. In this algorithm, we define each frequency bandwidth as 2 Hz and each temporal bandwidth as 1 second. Also, for frequency bands, the band start frequency increment is 1 Hz and for temporal bands, the start temporal increment is 0.5 second. Therefore, the frequency band is from 0-2 Hz, 1-3 Hz, 2-4 Hz, … to 28-30 Hz and temporal bandwidth is 0-1s, 0.5-1.5s, 1-2s, … to 3-4s. By combining the frequency and temporal bands, the aforementioned matrix is generated. In this matrix, an element with the most significant signal intensity (highest signal amplitude in the frequency domain) is selected as the optimal motor imagery element and fed into CSP feature extraction and classification algorithm.

### C. Common Spatial Pattern (CSP)

Originally, a common spatial pattern algorithm is designed for two-class feature extraction. The CSP objective is to transfer high dimensional EEG signals into a low dimension spatial subspace with a proper transformation matrix [21]. An optimal transformation matrix could separate two classes of data based on their variance, as shown in Figures 4a and 4b. According to these figures, the variance of class A is maximum in Feature 1 axis but minimum in Feature 2 axis while the variance of class B is opposite to class A. The details of the transformation matrix are illustrated below.

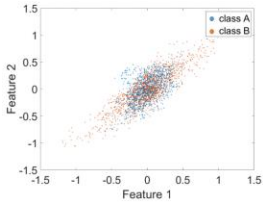 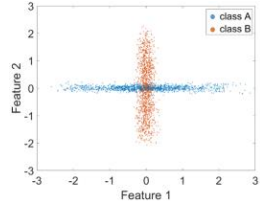

Figure 4a. EEG signal before CSP transformation  Figure 4b. EEG signal after CSP transformation

Assume a bi-class motor imagery experiment was conducted in $n$ trials each (left and right hand). $X_L$ and $X_R$ are denoted as left and right-hand motor imagery EEG signals for one trial, respectively. The dimension for both $X_L$ and $X_R$ is $N \times T$ where $N$ is the number of channels and $T$ is the number of samples. The covariance matrix ($N \times N$) for each class is

$$C_L = \frac{X_L X_L^T}{trace(X_L X_L^T)} \; ; C_R = \frac{X_R X_R^T}{trace(X_R X_R^T)} \quad (3)$$

where $X^T$ represents the transpose of $X$ and $trace(X)$ is the sum of the diagonal element of matrix $X$. By taking the mean value among all trails for each class, the average covariance matrix is

$$\overline{C_L} = \frac{1}{i} * \sum_1^i C_L^i \; ; \; \overline{C_R} = \frac{1}{i} * \sum_1^i C_R^i \quad (4)$$

where $\overline{C_L}$ and $\overline{C_R}$ are the average covariance matrix for the left and right classes, respectively. Thus, the composite spatial covariance matrix is

$$C = \overline{C_L} + \overline{C_R} \quad (5)$$

The composite spatial covariance matrix can be decomposed to

$$C = V \lambda V^T \quad (6)$$

where $V$ and $\lambda$ are the eigenvector and eigenvalue for the composite matrix. After decomposition, a whitening transformation matrix could be obtained as

$$P = \lambda^{-\frac{1}{2}} V^T \quad (7)$$

where P is the whitening transformation matrix with a dimension of $N \times N$. Hence, the average covariance matrices after applying whitening transform is

$$W_L = P \overline{C_L} P^T \; ; \; W_R = P \overline{C_R} P^T \quad (8)$$

where $P^T$ is the transpose matrix, $W_L$ and $W_R$ are the left and right whitening transformed matrices, respectively. Since $W_L$ and $W_R$ are composed of the same whitening transformation matrix, they share the same eigenvectors. Besides, the sum of the eigenvalue between $W_L$ and $W_R$ should be an identity matrix $I$, as shown below

$$W_L = U \lambda_L U^T; \; W_R = U \lambda_R U^T; \; \lambda_L + \lambda_R = I \quad (9)$$

where $U$ is the shared common eigenvector with a dimension of $N \times N$, and $\lambda_L$ and $\lambda_R$ are the eigenvalues for

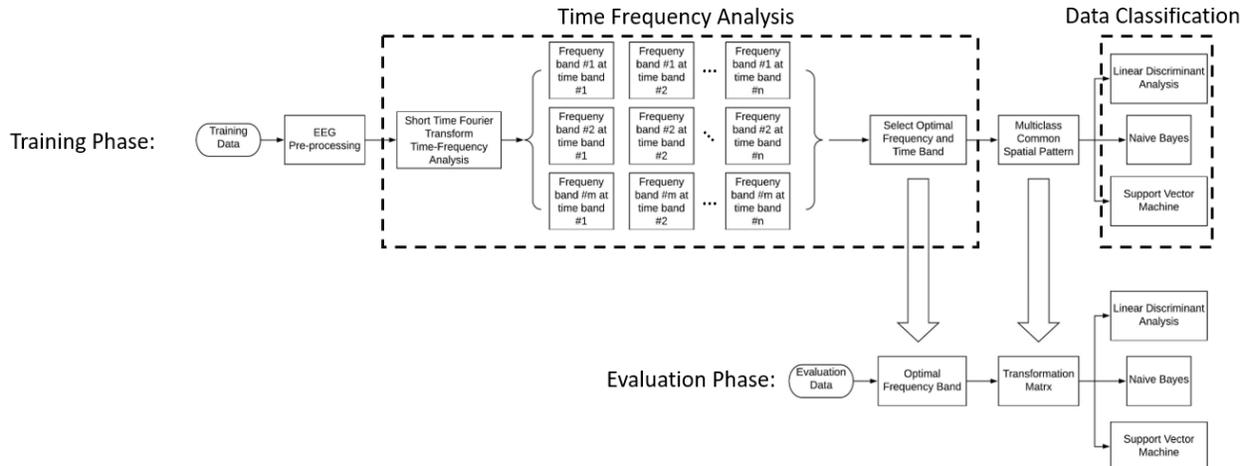

Figure 3. Proposed Motor Imageray Classification Flow Chart

left and right class, respectively. Finally, the CSP transformation is obtained by

$$Q' = U^T P \quad (10)$$

where Q is the CSP transformation matrix with the dimension $N \times N$. The original EEG signals were transformed as

$$Z = QX \quad (11)$$

where $Z$ is the CSP transformed results. In order to extend the two class CSP to M-CSP, joint approximate diagonalization (JAD) method was applied. M. Grosse-Wentrup and M. Buss provided a detailed discussion on how to extend a two class CSP to multiclass CSP [12].

## IV. RESULTS & DISCUSSIONS

The classification results are obtained based on BCI Competition IV Dataset 2a. The STFT time-frequency analysis and motor imagery classification results are interesting and presented below.

### A. STFT Time-Frequency Analysis

Figures 5a and 5b present time-frequency analysis results for the same subject with different numbers of trials in channel #8. Both figures show that time-frequency analysis successfully detects the specific time and frequency band in motor imagery events. As illustrated in the experiment protocol, subjects were given 3 seconds for motor imagery. However, it is not guaranteed that every subject could conduct motor imagery for all 3 seconds. Thus, the optimal temporal and frequency band is picked by motor imagery ERD/ERS behavior, where a place with higher signal intensity represents that motor imagery events occurred. For instance, Figure 5a shows that motor imagery is located at 10 Hz frequency from 2s to 2.5s. For Figure 5b, motor imagery still appears near 10 Hz frequency but it occurs at 1.5s to 2s after the cue started.

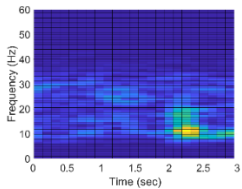
Figure 5a. Time-frequency plot for a subject in trial 25

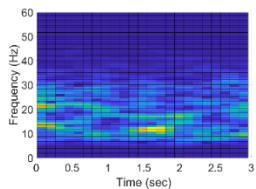
Figure 5b. Time-frequency plot for a subject in trial 15

According to these results, we found that for the same subject, the optimal frequency band is similar for every trial but the time band is different. The standard deviation is 1.2 for optimal frequency while it is 7.5 for time band. These results are reasonable because the ERD/ERS behavior is similar for the same subject but the time for the subject to start and stop motor imagery is varied in each trial. Therefore, in the proposed algorithm, an optimal frequency band is the average value among all trials while temporal band selection depends on each trial.

### B. Computation Time Results

The proposed TFCSP algorithm computation time is compared with FBCSP and traditional CSP (TDCSP) by using MATLAB timing function, as shown in Table 1. The proposed algorithm average calculation time is 37.22% faster than the FBCSP method. The proposed algorithm requires multi-class CSP feature extraction and classification calculation for one time, whereas the FBCSP requires nine times. Therefore, the proposed algorithm computation time is dramatically decreased. Moreover, since the proposed algorithm requires additional STFT analysis compared with the TDCSP, the TFCSP algorithm computation time is 4.98% longer than the TDCSP.

TABLE II. MATLAB COMPUTATION TIME COMPARISON BETWEEN FBCSP, TFCSP & TDCSP

|  | FBCSP (s) | TFCSP (s) | TDCSP (s) |
|---|---|---|---|
| Subject1 | 26.01 | 18.88 | 17.72 |
| Subject2 | 25.60 | 18.81 | 18.19 |
| Subject3 | 24.86 | 18.73 | 17.41 |
| Subject4 | 25.47 | 19.07 | 17.87 |
| Subject5 | 25.37 | 18.53 | 17.81 |
| Subject6 | 25.49 | 18.53 | 17.77 |
| Subject7 | 25.40 | 18.44 | 17.62 |
| Subject8 | 26.52 | 18.55 | 17.75 |
| Subject9 | 25.88 | 18.51 | 17.53 |
| **Average** | **25.62** | **18.67** | **17.74** |

### C. Motor Imagery Classification Results

The proposed algorithm classification results are compared with BCI Competition IV top three winner results, as shown in Table 2. With simplifying calculation procedures and analysis methods, the TFCSP algorithm overall detection accuracy is not as high as the algorithms with complex calculations (FBCSP), but, as aforementioned, the computation efficiency is increased.

TABLE I. KAPPA RESULTS COMPARISON BETWEEN PROPOSED METHODS & BCI COMPETITION IV RESULTS

|  | TFCSP Algorithm | | | BCI Competition Results | | |
|---|---|---|---|---|---|---|
| **Subjects** | **LDA** | **NVB** | **SVM** | **1st Place** | **2nd Place** | **3rd Place** |
| Subject1 | 0.64 | 0.47 | 0.62 | 0.68 | **0.69** | 0.38 |
| Subject2 | **0.43** | 0.34 | 0.36 | 0.42 | 0.34 | 0.18 |
| Subject3 | 0.71 | 0.49 | 0.76 | **0.75** | 0.71 | 0.48 |
| Subject4 | 0.38 | 0.25 | 0.40 | **0.48** | 0.44 | 0.33 |
| Subject5 | 0.29 | 0.14 | 0.29 | **0.40** | 0.16 | 0.08 |
| Subject6 | **0.39** | 0.24 | **0.39** | 0.27 | 0.21 | 0.14 |
| Subject7 | 0.63 | 0.53 | 0.59 | **0.77** | 0.66 | 0.29 |
| Subject8 | 0.57 | 0.26 | 0.57 | **0.75** | 0.73 | 0.49 |
| Subject9 | 0.61 | 0.50 | 0.62 | 0.61 | **0.69** | 0.44 |
| **Average** | **0.52** | 0.36 | 0.51 | **0.57** | 0.51 | 0.31 |

According to Table 2, kappa value is calculated and used for comparison with the top 3 results of the BCI competition IV. Kappa value is a statistical model evaluation method. The higher the kappa value, the more accurate the model is. The proposed algorithm analysis method kappa value is 39% better than the 3rd place in BCI competition IV, 0.2% better than the 2nd place, and 10% lower than the 1st place. However, the calculation procedures are much simpler than the 1st winner (FBCSP). In addition, by comparing different classifiers in the proposed algorithm, the LDA classifier exhibits the best classification performance, and the SVM classifier shows similar performance, which is only 0.75% lower than the LDA in kappa value. Even though the NVB classifier has the worst performance, it still outperforms the 3rd winner by 12.61% in kappa value.

## V. .Conclusions & Future Works

This paper proposed a TFCSP analysis method for multiclass motor imagery feature extraction with a more efficient computation. The performance of the proposed algorithm was examined using the experimental data provided by the BCI Competition open-source dataset. According to kappa values, the proposed algorithm classification accuracy is 39% higher, compared to the 3rd winner and slightly above the 2nd winner of the BCI Competition IV. A critical distinction is the reduced number of steps, and calculation time is 37.22% faster than the 1st winner in BCI Competition IV. This is substantially fewer calculations and hence computationally more efficient.

Currently, eight spatial features are selected from the CSP algorithm because of four motor imagery classes. In the future, multiple features could be selected. Moreover, because of its less computational load, the proposed method could be potentially applicable to online learning and conduct real-time motor imagery classifications.

References

[1] R. Abiri, S. Borhani, E. W. Sellers, Y. Jiang, and X. Zhao, "A comprehensive review of EEG-based brain–computer interface paradigms," *Journal of Neural Engineering,* vol. 16, no. 1, p. 011001, 2019/01/09 2019, doi: 10.1088/1741-2552/aaf12e.

[2] B. Graimann, B. Z. Allison, and G. Pfurtscheller, *Brain-Computer Interfaces: Revolutionizing Human-Computer Interaction*. Springer Publishing Company, Incorporated, 2013, p. 407.

[3] X. Guo, X. Wu, X. Gong, and L. Zhang, "Envelope detection based on online ICA algorithm and its application to motor imagery classification," in *2013 6th International IEEE/EMBS Conference on Neural Engineering (NER)*, 6-8 Nov. 2013 2013, pp. 1058-1061, doi: 10.1109/NER.2013.6696119.

[4] D. Cheng, Y. Liu, and L. Zhang, "Exploring Motor Imagery Eeg Patterns for Stroke Patients with Deep Neural Networks," in *2018 IEEE International Conference on Acoustics, Speech and Signal Processing (ICASSP)*, 15-20 April 2018 2018, pp. 2561-2565, doi: 10.1109/ICASSP.2018.8461525.

[5] T. Kaufmann, A. Herweg, and A. Kübler, "Toward brain-computer interface based wheelchair control utilizing tactually-evoked event-related potentials," *Journal of NeuroEngineering and Rehabilitation,* vol. 11, no. 1, p. 7, 2014/01/16 2014, doi: 10.1186/1743-0003-11-7.

[6] T. Mulder, "Motor imagery and action observation: cognitive tools for rehabilitation," (in eng), *Journal of neural transmission (Vienna, Austria : 1996),* vol. 114, no. 10, pp. 1265-1278, 2007, doi: 10.1007/s00702-007-0763-z.

[7] G. Pfurtscheller and F. H. Lopes da Silva, "Event-related EEG/MEG synchronization and desynchronization: basic principles," *Clinical Neurophysiology,* vol. 110, no. 11, pp. 1842-1857, 1999/11/01/ 1999, doi: https://doi.org/10.1016/S1388-2457(99)00141-8.

[8] R. Caldara, M.-P. Deiber, C. Andrey, C. M. Michel, G. Thut, and C.-A. Hauert, "Actual and mental motor preparation and execution: a spatiotemporal ERP study," *Experimental Brain Research,* journal article vol. 159, no. 3, pp. 389-399, December 01 2004, doi: 10.1007/s00221-004-2101-0.

[9] M. Sazgar and M. G. Young, "EEG Artifacts," in *Absolute Epilepsy and EEG Rotation Review: Essentials for Trainees*. Cham: Springer International Publishing, 2019, pp. 149-162.

[10] P. K. Saha, M. A. Rahman, and M. N. Mollah, "Frequency Domain Approach in CSP based Feature Extraction for EEG Signal Classification," in *2019 International Conference on Electrical, Computer and Communication Engineering (ECCE)*, 7-9 Feb. 2019 2019, pp. 1-6, doi: 10.1109/ECACE.2019.8679463.

[11] H. Ramoser, J. Muller-Gerking, and G. Pfurtscheller, "Optimal spatial filtering of single trial EEG during imagined hand movement," *IEEE Transactions on Rehabilitation Engineering,* vol. 8, no. 4, pp. 441-446, 2000, doi: 10.1109/86.895946.

[12] M. Grosse-Wentrup* and M. Buss, "Multiclass Common Spatial Patterns and Information Theoretic Feature Extraction," *IEEE Transactions on Biomedical Engineering,* vol. 55, no. 8, pp. 1991-2000, 2008, doi: 10.1109/TBME.2008.921154.

[13] A. Kai Keng, C. Zheng Yang, Z. Haihong, and G. Cuntai, "Filter Bank Common Spatial Pattern (FBCSP) in Brain-Computer Interface," in *2008 IEEE International Joint Conference on Neural Networks (IEEE World Congress on Computational Intelligence)*, 1-8 June 2008 2008, pp. 2390-2397, doi: 10.1109/IJCNN.2008.4634130.

[14] S. Selim, M. M. Tantawi, H. A. Shedeed, and A. Badr, "A CSP\AM-BA-SVM Approach for Motor Imagery BCI System," *IEEE Access,* vol. 6, pp. 49192-49208, 2018, doi: 10.1109/ACCESS.2018.2868178.

[15] A. P. Costa, J. S. Møller, H. K. Iversen, and S. Puthusserypady, "An adaptive CSP filter to investigate user independence in a 3-class MI-BCI paradigm," *Computers in Biology and Medicine,* vol. 103, pp. 24-33, 2018/12/01/ 2018, doi: https://doi.org/10.1016/j.compbiomed.2018.09.021.

[16] S. Sakhavi, C. Guan, and S. Yan, "Learning Temporal Information for Brain-Computer Interface Using Convolutional Neural Networks," *IEEE Transactions on Neural Networks and Learning Systems,* vol. 29, no. 11, pp. 5619-5629, 2018, doi: 10.1109/TNNLS.2018.2789927.

[17] C. Brunner, R. Leeb, G. R. Muller-Putz, A. Schlogl, and G. Pfurtscheller, "BCI Competition 2008 - Graz Data Set A," 2008.

[18] S. Bernhard and J. S. Alexander, "Support Vector Machines," in *Learning with Kernels: Support Vector Machines, Regularization, Optimization, and Beyond*: MITP, 2001, p. 1.

[19] D. W. T. H. R. T. Gareth James, *An introduction to statistical learning : with applications in R*. New York : Springer, [2013] ©2013, 2013.

[20] R. Polikar. "The Engineering Ultimate Guide to Wavelet Analysis: The Wavelet Tutorial." http://users.rowan.edu/~polikar/WTtutorial.html (accessed.

[21] M. Z. Baig, N. Aslam, and H. P. H. Shum, "Filtering techniques for channel selection in motor imagery EEG applications: a survey," *Artificial Intelligence Review,* journal article February 28 2019, doi: 10.1007/s10462-019-09694-8.